\documentclass[prl,twocolumn,a4paper,superscriptaddress,showpacs]{revtex4}

\usepackage{graphics}
\usepackage{ae}
\usepackage{aecompl}
\usepackage{hyperref}

\newcommand{\be}{\begin{equation}}
\newcommand{\ee}{\end{equation}}
\newcommand{\bea}{\begin{eqnarray}}
\newcommand{\eea}{\end{eqnarray}}
\newcommand{\bc}{\begin{center}}
\newcommand{\ec}{\end{center}}

\newcommand{\la}{\left<}
\newcommand{\ra}{\right>}

\begin{document}

\title{Long Range Bond-Bond Correlations in Dense Polymer Solutions}

\date{\today}

\author{J.P. Wittmer}
\email{jwittmer@ics.u-strasbg.fr}
\author{H. Meyer}
\author{J. Baschnagel}
\author{A. Johner}
\affiliation{Institut Charles Sadron, 6 Rue Boussingault, 67083 Strasbourg, France}
\author{S. Obukhov} 
\affiliation{Institut Charles Sadron, 6 Rue Boussingault, 67083 Strasbourg, France}
\affiliation{Department of Physics, University of Florida,
Gainesville FL 32611, USA}
\author{L.~Mattioni}
\affiliation{LPMCN, Universit\'e Claude Bernard \& CNRS, Lyon, France}
\author{M.~M\"uller}
\affiliation{Institut f\"ur Physik, Johannes Gutenberg-Universit\"at, 
Staudinger Weg 7, D-55099 Mainz, Germany}
\author{A.N. Semenov}
\affiliation{Institut Charles Sadron, 6 Rue Boussingault, 67083 Strasbourg, France}

\begin{abstract} 
The scaling of the bond-bond correlation function $P_1(s)$ along linear polymer
chains is investigated with respect to the curvilinear distance, $s$, along the 
flexible chain and the monomer density, $\rho$, via Monte Carlo and molecular 
dynamics simulations.
Surprisingly, the correlations in dense three dimensional solutions are found 
to decay with a power law $P_1(s) \sim s^{-\omega}$ with $\omega=3/2$
and the exponential behavior commonly assumed is clearly ruled out for long chains.
In semidilute solutions, the density dependent scaling of
$P_1(s) \approx g^{-\omega_0} \left(s/g\right)^{-\omega}$
with $\omega_0=2-2\nu=0.824$ ($\nu=0.588$ being Flory's exponent) is set by the
number of monomers $g(\rho)$ contained in an excluded volume blob of size $\xi$.
Our computational findings compare well with simple scaling arguments
and perturbation calculation. 
The power-law behavior is due to self-interactions of chains 
caused by the chain connectivity and the incompressibility of the melt.
This study suggests a careful reexamination of the operational 
definitions used for the experimental determination of the persistence length.
\end{abstract}

\maketitle

\begin{figure}[t]
\scalebox{0.35}{\includegraphics*{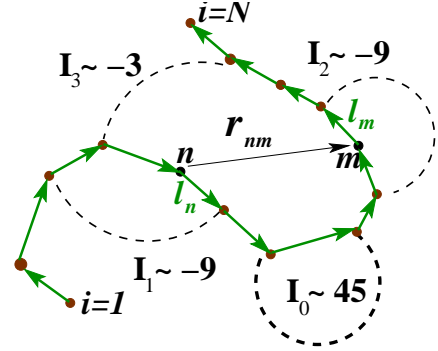}}
\vspace*{-0.3cm}
\caption{Sketch of the bond-bond correlation $\la l_n \cdot l_{m=n+s} \ra$ 
of the bond vectors $l_i=r_{i+1}-r_{i}$, $r_i$ being the position vector of
monomer $i$. 
The bond-bond correlation function is defined as the first Legendre polynomial
$P_1(s)= \la l_n \cdot l_{n+s} \ra_n/ l^2$ over all possible bond pairs with
curvilinear distance $s=m-n \ge 0$
normalized by the mean squared bond length $l^2=\la l_n^2 \ra$.
%
The typical size of the corresponding chain segments
is $R(s) = \la r_{nm}^2 \ra_n^{1/2}$ with $r_{nm}=r_{m=n+s}-r_n$. 
%
%
%
The dashed lines show the four relevant graphs of the analytical 
perturbation calculation.
The numerical factors indicate the relative weights of the leading 
$1/s^{3/2}$ contributions to $P_1(s)$.
\label{figsketch}
}
\end{figure}

%
%
In this Letter we study the correlations of the directions of bonds along polymer 
chains in semidilute solutions and melts \cite{Flory,degennesbook,DE86}.
%
We focus on flexible monodisperse chains of $N$ monomers (cf. Fig.~\ref{figsketch}) 
under good
solvent conditions in three dimensions ($d=3$) where both the bond length $l$
and the excluded volume screening length $\xi$ \cite{DE86,ME81} are always much
smaller than the chain end-to-end distance $R_e$.
%
%
In principle, once the bond-bond correlations are computed all other conformational 
single chain properties can be derived. Importantly, being the (second) 
derivative of the spatial distances along the chains, they allow us to probe 
directly --- without trivial ideal contributions --- the non-gaussian corrections 
proposed recently \cite{SJ03}.
As we shall see, these corrections are crucial to make the description
of dense polymer systems, first proposed by Flory \cite{Flory}
and later corroborated by Edwards \cite{DE86,ME81}, fully self-consistent. 
The bond-bond correlation function $P_1(s)$ is generally believed to decrease 
{\em exponentially} \cite{Flory}.
This belief is based on the few simple single chain models which have been 
solved rigorously \cite{Flory,Kratky} and on the assumption 
that {\em all} 
long range interactions are negligible on distances larger than $\xi$ due to the
screening mechanism described by Edwards \cite{DE86,ME81}. 
Hence, only correlations along the backbone of the chains are expected to matter and 
it is then straightforward to work out that an exponential cut-off is inevitable
due to the multiplicative loss of any information transferred recursively 
along the chain \cite{Flory}.
%

%
%
%
We demonstrate here that this assumption is in fact incorrect and that  
unexpected long range correlations remain.
They are responsible for a scale free {\em power law} regime with 
$P_1(s) = c_a(\rho) s^{-\omega}$ for $g(\rho) \ll s \ll N$ ($g(\rho)$ being 
the number of monomers per blob at monomer density $\rho$) characterized 
by an exponent $\omega > 1$ and a density dependent amplitude.
%
%
Our simulation results are presented first and discussed together with simple
scaling arguments. 
We focus on the scale free limit and finite chain size effects ($s \rightarrow N$) 
are considered more briefly. The analytical calculation 
(summarized graphically in Fig.~\ref{figsketch}) is outlined  
at the end of paper. 

\begin{figure}[t]
\scalebox{0.35}{\includegraphics*{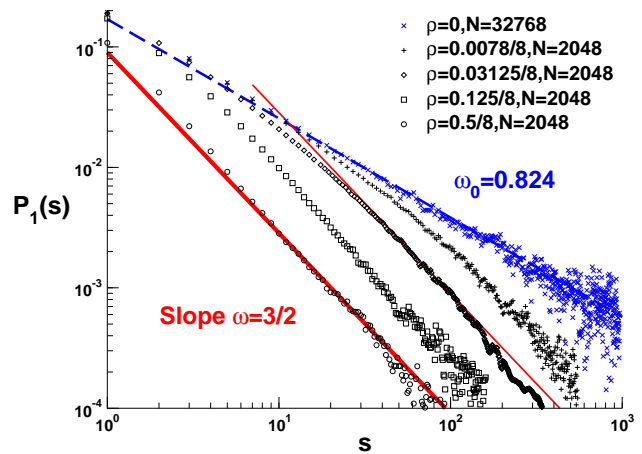}}
\vspace*{-0.3cm}
\caption{
The bond-bond correlation function $P_1(s)$ for BFM systems of different 
densities as indicated in the legend. 
The two lines indicate power laws corresponding to the asymptotic regime for 
dilute ($\omega_0 \approx 0.824$, dashed line) 
and dense ($\omega=3/2$, bold lines) solutions, respectively.
The observation of the second power law regime is the central result of this work. 
For systems in the so-called semidilute regime where
the excluded volume is sufficiently large both exponents can be seen. 
The curvilinear distance at the crossover corresponds to the number of 
monomers per blob $g(\rho)$.
}
\label{figliljphi}
\end{figure}

\begin{figure}[bt]
\scalebox{0.35}{\includegraphics*{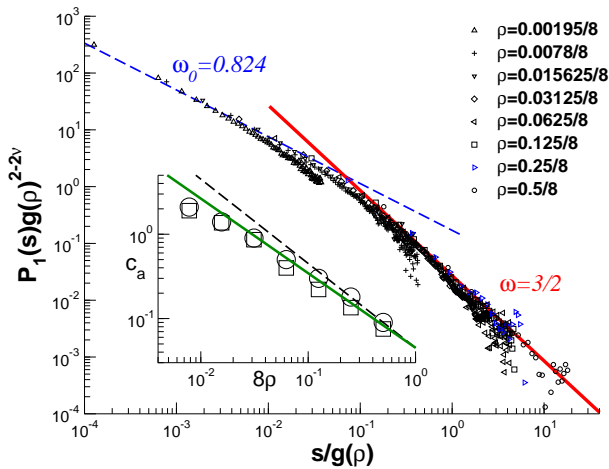}}
\caption{Crossover scaling for $P_1(s)$
for the BFM ($N=2048$) for different densities. 
In the main panel, the rescaled bond-bond correlation function
$f(x)=P_1(s) g(\rho)^{\omega_0}$ is plotted versus the natural
scaling variable $x=s/g(\rho)$ assuming 
$g(\rho) \propto \rho^{-1/(3\nu-1)}$.
Note that for large $s/N$ the final cut-off (see Eq.~(\ref{eqfull})) becomes visible.
In the inset we explicitly verify the density dependence of the 
amplitude $c_a(\rho)$ obtained from $P_1(s)$ (spheres) and $R(s)$ (squares)
using Eq.~(\ref{eqca}) and Eq.~(\ref{eqRnmdense}), respectively.
The prediction for the melt regime, Eq.~(\ref{eqca}), 
is indicated by the dashed line, 
the scaling 
$c_a \sim g(\rho)^{\omega-\omega_0}\sim \rho^{-0.885}$ 
for the semidilute regime by the bold line.
}
\label{figphiscal}
\end{figure}

\begin{figure}[ht]
\scalebox{0.35}{\includegraphics*{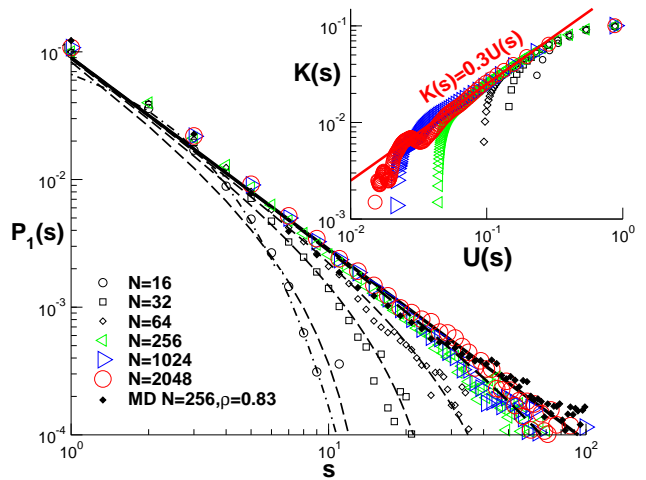}}
\caption{Bond-bond correlation function versus $s$ at melt densities
for different $N$ for the BFM ($\rho=0.5/8$) and
the bead-spring model ($\rho=0.83$) for $N=256$. 
The bold line indicates Eq.~(\ref{eqca}) with $c_a \approx 0.09$
for long BFM chains.
The dashed-dotted curve $P_1(s) \approx \exp(-s/1.5)$ shows that 
exponential behavior is compatible with the small $N$ data.
The dashed lines correspond to the complete theoretical 
prediction Eq.~(\ref{eqfull}) for finite chain lengths 
$N=16, 32, 64$ and $256$ (from left to right).
Considering that the theory does not allow for any free fitting parameter
the agreement is good.
In the inset we check directly the recursion relation
Eq.~(\ref{eqrecursion}) by plotting $K(s)$ 
versus $U(s)=s/R(s)^3\rho$
(same symbols as in main figure). The proportionality of $K(s)$ and $U(s)$
is well confirmed (bond line) by our data for $1 \ll s \ll N$.
\label{figliljmelt}}
\end{figure}

%
%
Two standard simulation methods for coarse-grained polymer chains \cite{BWM04} 
have been used to equilibrate long flexible polymer chains for 
which the bond-bond correlation functions presented in Figs.~\ref{figliljphi},
\ref{figphiscal} and \ref{figliljmelt} below have been computed. 
The body of our data comes from the ``bond fluctuation model" (BFM) --- 
a lattice Monte Carlo scheme where a monomer occupies $8$ lattice sites 
(i.e., the volume fraction is $8 \rho$) and the bonds $l_i$ between adjacent 
monomers can vary in length and direction, subject only to excluded volume 
constraints and entanglement restrictions \cite{BWM04,BFM}. 
Using a mixture of local, slithering snake and 
double-bridging moves 
\cite{BWM04,Auhl} we have created ensembles with chain lengths 
ranging up to $N=2048$ for densities between $\rho=0.00195/8$ 
to $\rho=0.5/8$ contained in large periodic boxes of linear size $L=512 \gg R_e$ 
which allows us to eliminate finite-size effects.
As can be seen in Fig.~\ref{figliljphi} we have also studied single 
chains up to $N=32768$ in an infinite non-periodic box ($\rho=0$)
to characterize properly the dilute reference point.
Additional molecular dynamics simulations of the bead-spring model discussed 
in \cite{BWM04} have been performed to check explicitly that our results are not 
caused by lattice artifacts. 
As shown in Fig.~\ref{figliljmelt} for one example with $N=256$ and $\rho=0.83$, 
identical behavior (although with reduced statistics) has been found.

%
%
Bond-bond correlation functions for different densities are presented in 
Fig.~\ref{figliljphi} where we focus first on the behavior of long chains 
($s \ll N)$. 
%
In the dilute limit ($\rho \rightarrow 0$)
the athermal BFM chains are well fitted by $R(s) = b_0 s^{\nu}$,
$\nu=0.588$ being Flory's good solvent exponent and $b_0\approx 3.0$. 
The power law slope with $\omega=\omega_0 \equiv 2-2\nu \approx 0.824$ 
and $c_a= \nu (2\nu-1) (b_0/l)^2$ observed in this limit (dashed line) is 
in fact expected \cite{omegadilute} from the  
formula
\begin{equation}
\la l_n \cdot l_m \ra =
\la \partial_n r_n \cdot \partial_m r_m \ra 
= -\frac{1}{2} \partial_n \partial_m \la r_{nm}^2 \ra
\label{eqRsPs}
\end{equation}
expressing $P_1(s)$ as the second derivative of $R(s)$.
It is crucial that $2\nu > 1$ and, hence, $\omega_0 < 1$, i.e. the integral
over the correlation function is dominated by its {\em upper} bound. Otherwise,
$\omega$ could not be governed by the exponent $\nu$ describing the asymptotic chain size.
Interestingly, it follows also directly from Eq.~(\ref{eqRsPs}) for ideal gaussian chains 
($2\nu\rightarrow 1$ : $c_a \rightarrow 0$) {\em without} additive power law corrections that 
$P_1(s)$ must decrease stronger than any power law, i.e. exponentially.

%
%
Coming back to Fig.~\ref{figliljphi}, we note that also for finite densities
$P_1(s)$ coincides with the dilute power law for small curvilinear distances $s$ 
where each chain segment interacts primarily with itself. At larger $s \gg g(\rho)$
where the chains overlap and form a ``melt of blobs" the correlation function
decreases much faster, however not exponentially as one might expect, but with a 
second power law regime with $\omega \approx 3/2$ (bold lines). 
For the BFM the density $\rho=0.5/8$ corresponds to a polymer melt since 
the excluded volume size becomes of order of the average bond size \cite{BWM04}. 
We observe indeed that the dilute slope disappears in the melt limit and the 
correlation function can be fitted (bottom bold line) over nearly two orders 
of magnitude by 
\begin{equation}
P_1(s)= c_a s^{-\omega} \text{ with } 
\omega=3/2 , c_a = \frac{\sqrt{6/\pi^3}}{4\rho l^3}
\label{eqca}
\end{equation} 
where we anticipate the analytical prediction for the asymptotic scale free limit.
Considering that all parameters are known 
the agreement is excellent although a slightly better fit could be obtained by 
accepting a larger exponent.  
Obviously, the chain statistics in the semidilute and melt regime must become  
in leading order gaussian, i.e. the integral over Eq.~(\ref{eqRsPs}) must
be dominated by the lower bound in agreement with the finding $\omega >1$.
In fact, Eq.~(\ref{eqca}) is consistent with
\begin{equation}
R(s)^2 = b^2 s - \frac{2 c_a l^2}{(\omega-1)(2-\omega)}  s^{2-\omega}
\label{eqRnmdense}
\end{equation}
with $b$ 
being the statistical segment length 
\cite{DE86,ME81}.
As long as $\omega <2$, the correction term (being due to an upper integration bound)
contributes a {\em negative}, but ever decreasing contribution to the chain distance, 
i.e., $R(s)^2/s$ approaches $b^2$ from below.
Interestingly, Eq.~(\ref{eqRnmdense}) shows that the computation of $P_1(s)$ 
allows to highlight directly corrections to gaussian behavior which can be 
blurred by the large ideal term if moments of spatial distances are considered. 
%

%
%

The scaling of $P_1(s)$ and $c_a$ with density 
are analyzed in Fig.~\ref{figphiscal}. Following the classical
density crossover scaling \cite{degennesbook} we take as natural scaling 
variable $x=s/g(\rho)$ and 
\begin{equation}
P_1(s) g(\rho)^{\omega_0} = f(x) \sim
\left\{ \begin{array}{ll}
x^{-\omega_0} & \mbox{, $x \ll 1$}\\ 
x^{-\omega} & \mbox{, $x \gg 1$}
\end{array}\right.
.
\label{eqphiscal}
\end{equation}
This is well borne out by the data.
To be specific, we have used  
$g(\rho) \propto \rho^{-1/(3\nu-1)}$ \cite{footprefactor}
although the power law form cannot hold strictly for large densities
where the semidilute blobs become too small and the assumed density 
dependence must be a crude estimate. 
Considering this, the scaling collapse for all densities
(obtained without any arbitrary shift parameter) is far from obvious
and quite satisfactory. 
%
%
The amplitude $c_a(\rho)$ of the non-gaussian corrections has been fitted 
directly from $P_1(s)$ and $R(s)$ assuming $\omega=3/2$. As can be seen in the inset, 
the data compare rather well with the prediction for the semidilute regime
$c_a \sim g(\rho)^{\omega-\omega_0}\sim \rho^{-0.885}$. 
%
 
%
%

In the following we concentrate on dense melts. 
In Fig.~\ref{figliljmelt} chain 
length effects are discussed at constant density. Data from the bead-spring 
model have also been included to demonstrate the universality of the result. 
To collapse this correlation function on the BFM data it has been vertically 
shifted by the ratio of the amplitudes $c_a$ calculated from Eq.~(\ref{eqca}) 
for BFM and bead-spring model. 
As can be seen for $N=16$, exponentials are compatible with the data of
short chains. (This might explain how the power law scaling 
has been overlooked in previous numerical studies since good statistics
for large chains ($N > 1000$) has only become available recently.)
However, it is clearly shown that $P_1(s)$ approaches systematically the 
scale free asymptote with increasing $N$. The departure from
this limit is fully accounted for by the theory if chain end effects
are carefully considered (dashed lines). 
Generalizing Eq.~(\ref{eqca}), perturbation theory yields 
\begin{equation}
P_1(s) = c_a s^{-3/2} a(x) + c_b s^{-1/2} b(x)
\label{eqfull}
\end{equation}
where we have set $x=\sqrt{s/N}$ and $c_b=4 c_a/N$ and defined the
functions $a(x) = b(x) (1+3x+x^2)$ and $b(x) = (1-x)^2/(1+x)$.
For $x \ll 1$ only the first term contributes:
however, as $s \rightarrow N$  both terms become 
of similar magnitude. 
Both contributions to the correlation function vanish
rigorously in this limit, and $P_1(s) \approx (1 - x)^2$.
%

%
We outline now very briefly how Eqs.~(\ref{eqca}) and 
(\ref{eqfull}) have been obtained using perturbation calculation 
\cite{footfield}. Closely following Edwards \cite{DE86} we first determine 
$\la r_{nm}^2 \ra \approx \la r_{nm}^2 \ra_0 (1+ \la U \ra_0)-\la r_{nm}^2 U \ra_0$  
and then, using Eq.~(\ref{eqRsPs}), the bond-bond correlations. 
Here, $\la ... \ra_0$ denotes the average for the distribution function 
of the unperturbed ideal chain and $U = \int_0^N dk \, \int_0^k dl \, v_1(r_{kl})$
the effective perturbation potential. 
The simplest approximation \cite{DE86} for the pair interaction potential 
in Fourier space is
$v_1(q)= v q^2/(q^2+\xi^{-2})$
with $v$ being the bare excluded volume parameter and $\xi=(l^2/12\rho v)^{1/2}$
the screening length.
This is sufficient for calculating the scale free regime 
(Eq.~(\ref{eqca})). The graphs which contribute to $\la l_n \cdot l_m \ra$ and 
their relative weights are indicated in Fig.~\ref{figsketch}. 
Note that the interactions described by the strongest graph $I_0$ align the bonds 
$l_n$ and $l_m$ while the others tend to reduce the effect.  
More care is needed to describe properly the finite chain size corrections
Eq.~(\ref{eqfull}), and the Pad\'e approximation for the 
intrachain structure factor must be used here.
%
It can be readily shown that the different integrals are always dominated by 
long wavelength physics where $v_1(q)\rightarrow q^2 l^2/12\rho \sim v^0$. 
Therefore, the coefficients
$c_a$ and $c_b$ of Eqs.~(\ref{eqca}) and (\ref{eqfull}) do not depend on
local interaction parameters, such like $v$
\cite{footcoarsegraining}.

%
%
Since these calculations are lengthy  
we present a simple scaling argument
for the $\omega$ exponent which captures the central physical idea \cite{SJ03}. 
As suggested by Eq.~(\ref{eqRnmdense}) a direct measure for the non-gaussian 
corrections can be defined by 
$K \equiv (R^2(2s)- 2 R^2(s))/(2R^2(s)) \approx c_a s^{1-\omega}$
where we compare the size of a segment of length $2s$ with the size of two 
segments of length $s$ joint together. 
Equivalently, this can be read as a measure for the swelling of a chain 
where initially the interaction energy $U$ between the two halves has 
been switched off. Obviously, $K$ must be a functional of $U$ and $K(U=0)=0$.
Perturbation theory tells us that the lowest order of the expansion of 
$K$ in $U$ must be linear. 
What is still missing is an estimate for the scaling of the interaction
energy between the segments. For {\em incompressible} melts  
it is well known \cite{degennesbook} that the energy penalty for joining two 
chains of length $N$ is set by the depth of the correlation hole, 
$U(N) \approx N/R(N)^d/\rho \sim N^{1-d\nu}$ \cite{SJ03,footUN}. 
This generalizes to arbitrary $s$. 
Hence, in $d=3$ the interaction energy is small ($U(s) \sim 1/s^{1/2}$) 
which finally justifies the perturbation calculation.
Taking everything together we obtain  $\omega = d \nu = 3/2$ and the 
important recursion relation
\begin{equation}
K(s) = \frac{R^2(2s)- 2 R^2(s)}{2R^2(s)} \approx U(s) \equiv \frac{s/R(s)^d}{\rho}.
\label{eqrecursion}
\end{equation}
The recursion relation provides a very compact and entirely self-consistent
description of dense polymer melts ($d>2$). 
To lowest order it leads to Eqs.~(\ref{eqca}) and (\ref{eqRnmdense}).
This shows that while the chain configurations must be gaussian
to leading order, corrections do  necessarily occur ---
even for very long chains. 
Eq.~(\ref{eqrecursion}) has been directly validated
in the inset of Fig.~\ref{figliljmelt} where we have plotted $K(s)$ 
versus $U(s)$. The predicted linearity is well confirmed for
large segments ($s \gg 1$). 
The recursion relation being independent
of the total chain length $N$ does not capture the scaling for very large segments.
Hence, the breakdown of the scale free behavior for $K(U)$ (inset) and $P_1(s)$ 
(main figure) are both due to the same finite chain size effect. 
%
%

%
%
In summary, we have shown that long range correlations exist
in polymer melts and semidilute solutions due to both chain connectivity and 
incompressibility. 
Their most striking effect is the power law 
asymptote for the bond-bond correlation function 
allowing a direct numerical test 
of the non-gaussian corrections suggested by the self-consistent 
recursion relation Eq.~(\ref{eqrecursion}).
An important consequence of this work arises for an experimentally
relevant quantity, the static structure factor $S(q)$.
In fact, simulation and theory show distinct non-monotoneous behavior of
$q^2 S(q)$ {\em vs.} $q$ (Kratky plot) due to the non-gaussian corrections. 
This suggests a possible route for experimental verification and 
is cause of serious concerns with regard to the standard operational 
definition and measure of the persistence length from the assumed Kratky plateau.
Finally, we point out that the physical mechanism which has been sketched 
above is rather general and should not be altered by details such as a 
finite persistence length 
--- at least not as long as nematic ordering remains negligible. 
This is in fact confirmed by preliminary and on-going simulations.
 
\begin{acknowledgments}
We thank J.-L. Barrat and K.~Binder for 
useful discussions and acknowledge financial support from the 
LEA and the ESF SUPERNET programme as well as computational 
resources by IDRIS, Orsay. 
\end{acknowledgments}


\begin{thebibliography}{99}
\bibitem{footUN}
Interestingly, $U$ does not depend explicitly on the excluded volume parameter $v$.
The scaling of $U(r,N)$ has been verified directly by measuring the pair correlation 
function $p(r)$ of the chain center of masses and using 
$\exp(-U(r,N)) \approx p(r)/p(\infty)$.
\bibitem{footprefactor}
The prefactor has been obtained from the scaling of $R(s)$ 
for $\rho < 0.125/8$.
\bibitem{footfield}
A similar description has been also obtained for polydisperse living
polymers using the polymer-magnetic analogon. This will be presented elsewhere.
\bibitem{footcoarsegraining}
Contrary to this, local interaction parameters do modify the statistical
segment length $b$ \cite{DE86,ME81}.
%
Interestingly, the generic behavior of the bond-bond-correlation function
does not change under systematic coarse-graining $\lambda$ neighboring bonds. 
Our simulation results confirm again the expected scaling 
$P_1(s;\lambda) \approx
\frac{1}{(\rho/\lambda) (l\lambda^{1/2})^{3}} (\lambda/s)^{\omega=3/2}$
for long chains.
\end{thebibliography}
\end{document}